
\magnification = 1200
\tolerance 1000
\pretolerance 1000
\baselineskip 20pt
\overfullrule = 0pt
\hsize=16.3truecm
\def\sp{\kern 1em}

\centerline {\bf COHERENT TRANSPORT THROUGH A QUANTUM TROMBONE. }
\vskip 0.5truecm

\centerline {\bf  N.K. Allsopp and C.J. Lambert, }

\centerline {\bf School of Physics and Materials}

\centerline {\bf Lancaster University}

\centerline {\bf Lancaster   LA1 4YB}

\centerline {\bf U.K.}

\vskip 1.00truecm

\noindent
{\bf ABSTRACT.}

We highlight a new \lq\lq transverse" interference effect, arising from the
coupling between electrons and holes,
induced by a superconducting island in contact
with a normal metal. As an example
we compute the  electrical conductance $G$ of
a T-shaped mesoscopic sample,
formed by joining a horizontal bar of normal metal to
a vertical leg of the same material. With a superconducting island
located on the vertical
leg and the current flowing horizontally, we examine the dependence
 of $G$ on the
distance $L$ of the island from the horizontal bar.
 Of particular interest is the differential conductance $G(E)$
at an applied voltage $E=eV$, which due to quantum interference
between electrons and Andreev reflected holes, is predicted to oscillate
with both $L$ and $E$. For a system with a spherical Fermi surface,
the period of oscillation with $E$ is $\pi E_F/k_FL$
and with $L$ is $\pi E_F/k_F E$, where $E_F$ is the Fermi energy
and $k_F$ the Fermi wavevector.


\vfil\eject
\vskip 0.5cm
\noindent {\bf 1. Introduction.}
\vskip 0.3cm
Mesoscopic superconductors, formed by adding one or more superconducting
islands to a phase coherent normal host, represent a new class of
quantum systems, whose realisation has been made possible by
recent advances in nano-lithographic techniques[1,2,3].
The possibility of such structures has led
many theorists to consider incorporating
superconductivity into modern theories of phase coherent transport[4-13]
and several experimental groups to carry out related investigations[14-18].
In this paper, we analyze a simple demonstration experiment, carried out
originally to address the prediction of superconductivity induced conductance
suppression[12,13]. The experiment[18] measures the
 electrical conductance $G$ of
a T-shaped sample of size less than the quasi-particle phase breaking length,
formed by joining a horizontal bar of silver to
a vertical leg of the same material. With a superconducting island
located on the vertical
leg, a distance $L$ from the T-junction
 and the current flowing horizontally,  the dependence of $G$ on the
 distance $L$ and on the
 applied magnetic field was measured.
 As the applied field was lowered and the island allowed to become
 superconducting,
macroscopic conductance changes of arbitrary sign were obtained,
 in agreement with
the predictions of references[12,13].

In what follows, we compute the conductance $G(eV) = \partial I/\partial V$
of such a T-shaped \lq\lq trombone" as the superconducting island is
slid along the vertical leg. To our knowledge, this quantity has not yet been
measured for such structures,
although the work of references [14-18] demonstrates that $G(eV)$
is experimentally accessible.
 We note that at zero temperature, in the strict linear response
 limit of vanishing applied voltage, the conductance oscillates with
 $L$ on the scale of the Fermi wavelength $\lambda_F$. This rapid variation
arises from normal reflection at the interface with the island
 and is found in quantum modulated transistors[19,20,21],
with no superconductivity.
In contrast, we predict that
the differential conductance at finite voltage $V$
is sensitive to quantum interference
between electrons and Andreev reflected holes and  show that
this  oscillates
with $L$ on the much longer scale
$\sim \pi E_F/(E k_F)$,
where $E=eV$,
$E_F$ is the Fermi energy and $k_F$ the Fermi wavevector.
Such an effect is reminiscent of Tomasch oscillations in tunnel
 junctions[22,23],
although
in the present context, the current flow is transverse to the direction
of the geometric feature causing interference.

\vskip 0.5cm
\noindent {\bf 2. Calculation of the electrical conductance.}
\vskip 0.3cm

To demonstrate this new interference effect,
 one notes that if the ends of the horizontal bar of the trombone are
 connected to  external reservoirs by normal, current carrying leads,
then the two probe, differential electrical conductance
is given by
$$G(E)=e[\partial I/\partial (\mu_1 - \mu_2)]_{(\mu_1-\mu_2=E)},$$
where $\mu_1,\, \mu_2$ are the chemical potentials of the
reservoirs.
At zero temperature, the analysis of references [7,8],
yields, in units of $2e^2/h$,

$$G(E)= {{1 - R_{pp}(E) + R_{hp}(E) + \alpha(T_{hh'}(E) - T_{ph'}(E))}
\over {1 + \alpha}},
\eqno{(1)}$$
where
$$\alpha = {{R_{hp}(E) + T_{h'p}(E)}\over{R_{p'h'}(E) + T_{ph'}(E)}}$$
and the notation of reference[7] has been adopted.
The coefficients $R_{pp}(E), R_{hp}(E)$
($R_{h'h'}(E), R_{p'h'}(E)$) are
probabilities for normal and Andreev
reflection respectively for quasi-particles (quasi-holes) from the left
(right) reservoir, while
$T_{p'p}(E), T_{h'p}(E)$ ($T_{hh'}(E), T_{ph'}(E)$) are the corresponding
transmission probabilities. These are obtained by tracing
over combinations of sub-matrices of the scattering matrix of the
structure[7,8].

Consider first the T-shaped structure shown in figure 1, which comprises
three one-dimensional wires, connected at a node. In the absence
of superconductivity, the Bogoliubov - de Gennes equation
$$
\left(\matrix{H_0(\underline r)
& \Delta(\underline r) \cr
\Delta^*(\underline r)
& -H_0^*(\underline r)\,}\right)
\left(\matrix{\psi(\underline r)\cr \phi(\underline r)}
\right) = E \left(\matrix{\psi(\underline r)\cr \phi(\underline r)}
\right) \eqno{(2)},$$
reduces to two separate equations for particles and holes.
For clean
wires, in a region occupied by the $j$th wire, a solution of the particle
(hole) equation
consists of incoming and outgoing plane waves of amplitudes
$v_k^{-1/2}A_j$, $v_k^{-1/2}B_j$
($v_q^{-1/2}\bar A_j$, $v_q^{-1/2}\bar B_j$) respectively, where
$v_k$ $(v_q)$ is the group velocity of a particle (hole) of energy
$E=\hbar^2k^2/2m - E_F = E_F -\hbar^2q^2/2m$. These satisfy
equations of the form

$$\left(\matrix{B_1\cr B_2\cr B_3 }\right) = S(E)
\left(\matrix{A_1\cr A_2\cr A_3 }\right) \eqno{(3)}$$
and
$$\left(\matrix{\bar B_1\cr \bar B_2\cr \bar B_3 }\right) = S^*(-E)
\left(\matrix{\bar A_1\cr \bar A_2\cr \bar A_3 }\right) \eqno{(4)}.$$

To investigate the possibility of geometric resonances,
we now introduce a superconducting boundary, a distance $L$
along the vertical leg of the structure. For convenience, we
consider an ideal boundary,
with negligible normal scattering,
which induces coupling between particles and holes only.
If $r_a(E)$ ($\bar r_a(E)$) are Andreev reflection coefficients for particles
(holes) incident on the interface, then the coupling takes the form
$$\bar A_3 = r_a(E) B_3 \,\,\,\,\,\,\,\,\,\,\, A_3=\bar r_a(E) \bar B_3
\eqno{(5)}.$$
This pair of equations allows us to eliminate the amplitudes
$A_3$, $\bar A_3$ from equations (3) and (4), to yield
an equation of the form (see appendix)
$$\left(\matrix{B_1\cr B_2\cr \bar B_1\cr \bar B_2 }\right)=
\left(\matrix{S^{pp}(E) & S^{ph}(E)\cr S^{hp}(E) & S^{hh}(E) }\right)
\left(\matrix{A_1\cr A_2\cr \bar A_1\cr \bar A_2 }\right) \eqno{(6)},$$
where the 2x2 sub-matrices $S^{ij}(E)$ satisfy the following
particle-hole symmetry relations
$S^{hh}(-E)=[S^{pp}(E)]^*$
and $S^{ph}(E)=-[S^{hp}(-E)]^*$.
All reflection and transmission coefficients are
obtained from the modulus squared of elements of these sub-matrices.
In particular,
$T_{p'p}(E)=\vert S_{21}^{pp}\vert^2$, $R_{hp}(E)=\vert S_{11}^{hp}\vert^2$,
$T_{hh'}(E)=\vert S_{12}^{hh}\vert^2$ and $R_{p'h'}(E)=\vert
S_{22}^{ph}\vert^2$.

As an example, consider a spatially symmetric scatterer, for which
$$S(E) = \left(\matrix {r&t&t\cr t&r&t\cr t&t&r}\right). \eqno{(7)}$$
In this case, the conduction formula (1) reduces to
$G(E)=(R_{ph}(E)+R_{h'p'}(E) + T_{hh'}(E)+T_{p'p}(E))/2$.
and in the limit $E/E_F << 1$, one obtains (see appendix)
$$T_{p'p}(E) \simeq \vert t \vert^2\vert 1+r^*t/d\vert^2 \eqno{(8)},$$
$$T_{hh'}(E) \simeq\vert t \vert^2\vert 1+rt^*/d\vert^2 \eqno{(9)},$$
$$R_{hp}\simeq R_{p'h'}\simeq \vert t\vert^4/\vert d\vert^2 \eqno{(10)},$$
where
$$d\simeq -\vert r\vert^2 - {\rm exp}[-2i(k-q)L]\eqno{(11)}.$$
In the above expressions,
only the denominator $d$
depends on the length $L$. At zero energy,
since $k-q \simeq k_F E/E_F$, this dependence vanishes.
At finite energy,  there is a slow
 oscillation with period given by
$$ k_F L E/E_F =  \pi \eqno{(12)}.$$

\vskip 0.5cm
\noindent {\bf 3. Numerical results in two-dimensions.}
\vskip 0.3cm

The above analysis demonstrates
 that transverse geometric oscillations are a generic
feature of structures of the kind shown in figure 1. For a more realistic
 interface,  which induces both normal and Andreev reflection,
it is clear from the analysis of references[19-21] that
there would be an additional, more rapid oscillation on the scale of $k_F L$.
To illustrate this feature and to show that slow oscillations survive
in more realistic systems of finite width, we now present the results of
a numerical simulation of the 2-dimensional T-shaped structure, shown in
figure 2 and described by a tight binding
Bogoliubov - de Gennes operator of the form
$$H =
\left(\matrix{H_0
& \Delta \cr
\Delta^*
& -H_0^*\,}\right) \eqno{(13)}.$$
In this equation $H_0$ a nearest neighbour Anderson model
on a square lattice, with off-diagonal hopping elements of magnitude unity
and $\Delta$ a diagonal matrix with on site, particle-hole
 couplings. The latter are chosen to
vanish in the normal regions and to have a value $\Delta_0$ in the region
occupied by the superconducting island (shown black in figure 2).
 The scattering region is chosen to be  $M=30$ sites wide and 6
 sites long and is connected to
external leads of width $M$.
Within the leads, the diagonal elements are
 equal to $\epsilon_0$ and within the insulating portion of the scattering
region (shown shaded)
the diagonal elements are set to an arbitrarily large value
$\bar\epsilon$.
 Within the remaining (conducting) portion of
the scattering region,
diagonal elements of $H_0$
are chosen to be random numbers, uniformly distributed
between $\epsilon_0 -W$ and $\epsilon_0+W$.
In what follows, the choice $\epsilon_0= 0$, $\bar\epsilon = 50$,
$\gamma=1$ was made and free end boundary conditions employed.
 In the experiments of reference[18], the conductance
per channel was estimated to be of order 0.1 $e^2/h$, which for the
system of figure 2, is obtained with the choice $W\simeq 0.1$.
The results of figure 3 were obtained with this choice of $W$ and
$\Delta_0=0.2$.

As well as possible oscillatory dependencies on $E$ and $L$,
it is of interest to examine the
effect of a magnetic field. Therefore
for each set of random diagonal elements, the conductance $G(E)$
was first obtained with zero magnetic field, then recomputed with
a finite flux $\phi$ through the conducting region.
For each choice of $E$ and $L$, the average conductance
$G(E,L)$ was computed from results obtained for 100
different realisations of the random diagonal elements, with the same  sets
of random numbers being used for each $E$ and $L$.
The squares in figure 3 show results for $\phi=0$, while the circles
show results for $\phi = 6 \phi_0$, where $\phi_0=hc/e$. For the latter,
the field was introduced through a Peierls substitution in the off-diagonal
elements of $H_0$.

At zero energy, the conductance oscillates on the scale of the lattice spacing,
which for the system of figure 2 is of order the Fermi wavelength $\lambda_F$.
At this energy, particle-hole symmetry ensures that there is no interference
between electrons and Andreev reflected holes from the normal-superconducting
interface. At finite energies, the wavelengths of particle and holes differ
by an amount proportional to $E$ and therefore quantum interference on
 a length scale
proportional to $E^{-1}$ can occur. This new oscillatory
 effect, is  distinct from more
familar Tomasch[22] and Macmillan-Rowell[23] oscillations, since the
latter occur in tunnel junctions, while in the structure
 considered here, there is no insulating barrier. In addition, the
 effect reported here occurs for $E <\Delta_0$ as well as $E>\Delta_0$.

 To obtain a crude estimate of the periods of oscillation shown in figure 3,
 one notes that
in the absence of disorder and superconductivity,
a vertical, normal lattice of width two sites
possesses a dispersion relation for particles of the form
$E=-2\gamma[{\rm cos}k_ya + {\rm cos}k^n_xa]$ and for holes,
$E=+2\gamma[{\rm cos}q_ya + {\rm cos}k^n_xa]$, where $k_x^na=n\pi/3,\,\,\,
n=1,2$.
Choosing $E=0$,  yields values for the allowed wavevectors $k_F^n$ on the
Fermi surface given by $k_F^na = \pi/3$ or $3\pi/3$. Hence for small $E$,
$E\simeq \gamma{\rm sin}k_F^na\,(k_y-q_y)a = \gamma (k_x-q_x)a\sqrt{3}/2$.
and the period of oscillation obtained from equation
(12) is given by
$$(L/a)(E/\gamma) \simeq 2\pi/\sqrt{3}\eqno{(14)}.$$
While this is a somewhat crude estimate, one notes that
for $E/\gamma = 0.2$, this yields $L\simeq 18$, which is of the same order
as the period shown in the bottom graph of figure 3.

Figure 3 shows that oscillations occur both in zero and finite magnetic
fields and as a consequence,
the
magneto-conductance can change sign, if either $L$ or $E$ are varied.
The oscillations predicted by figure 3 are a finite fraction of
the total conductance and therefore should be observable experimentally.
In practice it is not possibly to slide the superconducting island of a
quantum trombone, without growing a separate sample with an independent
set of random impurities. Consequently oscillations on the scale of
$\lambda_F$ are difficult to observe
and would be measured as random sample to sample fluctuations. In contrast
the longer scale variation in the differential conductance should be
accessible experimentally. Indeed, by fixing the length $L$
 and varying $E$,
the oscillations reported above should be observable in a single sample.
For $L=11$,
this behaviour is illustrated in figure 4, which shows a slow variation
with $E$ on a scale proportional to $L^{-1}$. For $L=11$, equation (14)
yields a the period of $E/\gamma \simeq 0.3$, compared
with a value $E/\gamma \simeq 0.2$ obtained from figure 4.

\vskip 0.5cm
\noindent {\bf 4. Discussion.}
\vskip 0.3cm
We have demonstrated that a superconducting island in contact with a normal
mesoscopic conductor, produces geometric resonances,
even if the superconductor is far removed from the
current path. Such effects should be observable in the mesoscopic structures
described in reference [18].
The above results, which are based on the assumption of a rigid
order parameter, do not explain a weak magnetic field feature found in [18].
Figures 3 and 4 also show that the magneto-conductance can change sign with
varying $E$ and $L$, but the magnitude of the effect is much smaller than
that observed in reference [18]. Since the theory of
reference[12] predicts that the electrical conductance of a hybrid
superconducting structure is sensitive to small changes in the magnitude
of the superconducting order parameter, a possible
explanation of the measured small field feature is that weak proximity tails
present in the silver are modified by the applied field. To avoid such tails,
it would be of interest to repeat these
experiments, but with a magnetic normal host such as
Nickel[24]

\vfil\eject
\noindent
$\underline {\hbox{\bf Appendix.}}$
\vskip 0.5cm
Combining equations (3) and (5) yields
$B_3=S_{33}A_3 + S_{31}A_1 + S_{32}A_2 = \bar A_3/r_a(E)$.
Similarly
combining equations (4) and (5) yields
$\bar B_3=\bar S_{33}\bar A_3 + \bar S_{31}\bar A_1 + \bar S_{32}
\bar A_2 =  A_3/\bar r_a(E)$,
where the $i,j$th element of $S_{ij}(E)$ has been
denoted $S_{ij}$ and the $i,j$th element of $S^*_{ij}(-E)$
denoted $\bar S_{ij}$. These yield expressions for
 $A_3$, $\bar A_3$ of the form
$$A_3= \alpha_1A_1 + \alpha_2A_2 + \beta_1\bar A_1 + \beta_2\bar A_2$$
and
$$\bar A_3= \bar\beta_1A_1 + \bar\beta_2A_2 + \bar\alpha_1\bar A_1
+\bar\alpha_2\bar A_2 ,$$
where

$\alpha_1=\bar S_{33}S_{31}/d$, $\alpha_2=\bar S_{33}S_{32}/d$,

$\bar\alpha_1= S_{33}\bar S_{31}/d$, $\bar \alpha_2= S_{33}\bar S_{32}/d$,

$\beta_1=\bar S_{31}/(d\,r_a(E))$, $\beta_2=\bar S_{32}/(d\,r_a(E))$,

$\bar\beta_1= S_{31}/(d\,\bar r_a(E))$, $\bar\beta_2= S_{32}/(d\,\bar r_a(E))$,
\noindent
with  $d=(r_a(E) \bar r_a(E))^{-1}-S_{33}\bar S_{33}$.

Substituting these back into equations (3) and (4) yields equation (6),
with
$$ S^{pp}=\left(\matrix{S_{11}+S_{13}\alpha_1 & S_{12}+S_{13}\alpha_2\cr
S_{21}+S_{23}\alpha_1 & S_{22}+S_{23}\alpha_2}\right)$$

$$ S^{ph}=\left(\matrix{S_{13}\beta_1 & S_{13}\beta_2\cr
S_{23}\beta_1 & S_{23}\beta_2}\right)$$

$$ S^{hh}=\left(\matrix{\bar S_{11}+\bar S_{13}\bar \alpha_1 &
\bar S_{12}+\bar S_{13}\bar \alpha_2\cr
\bar S_{21}+\bar S_{23}\bar \alpha_1 &
\bar S_{22}+\bar S_{23}\bar\alpha_2 }\right)$$

$$ S^{hp}=\left(\matrix{\bar S_{13}\bar \beta_1 & \bar S_{13}\bar \beta_2\cr
\bar S_{23}\bar \beta_1 & \bar S_{23}\bar \beta_2}\right)$$

For a N-S interface at a point $x_0$ along the
$x$-axis, Andreev's approximation
for reflection amplitudes is
$r_a(E)={\rm exp}i[-\theta-\phi+(k-q)x_0]$,
$\bar r_a(E)={\rm exp}i[-\theta+\phi+(k-q)x_0]$,
where $\phi$ is the phase of the superconducting order parameter
and
$\theta = {\rm tan}^{-1}[(\vert \Delta\vert^2-E^2)^{1/2}/E] +\pi\Theta(-E)$.
In view of particle-hole symmetry
$r_a(E)=-\bar r_a^*(-E)$, as expected.

For a symmetric scatterer, with  S-matrices of the form
$$S(E) = \left(\matrix {r&t&t\cr t&r&t\cr t&t&r}\right) , \,\,\,\,\,\,\,\,
\,\,\,
\,\,\,\,\,\,
\bar S(E) = S^*(-E) =\left(\matrix {\bar r&\bar t&\bar t\cr \bar t&
\bar r&\bar t\cr \bar t&\bar t&\bar r}\right) ,$$
one obtains
$$T_{p'p}(E)=\vert t\vert^2\vert 1+\alpha_1\vert^2 ,$$
$$T_{hh'}(E)=\vert \bar t\vert^2\vert 1+\bar\alpha_2\vert^2 ,$$
$$R_{hp}(E)= \vert \bar t\vert^2\vert\bar\beta_1\vert^2$$
and
$$R_{p'h'}(E)= \vert t\vert^2\vert\beta_2\vert^2 .$$
In these expressions,
$$\alpha_1=\bar t/d \hsize 1cm \bar\alpha_2=r\bar t/d$$
$$\bar\beta_1=t/(d\bar r_a) \hsize 1cm \beta_2=\bar t/(dr_a)$$
and
$$d=(r_a(E)\bar r_a(E))^{-1}-\vert r \vert^2 .$$

For small $E$, provided $t$, $r$ are slowly varying compared with
$(E/E_F)k_FL$, one may evaluate $r$, $t$ at $E=0$ and write
 $\bar t \simeq t^*$ and $\bar r \simeq r^*$.
This yields equations (8) to (11) of the main text.

\vfil\eject

\vskip 0.5truecm
\noindent
$\underline {\hbox{\bf Acknowledgements.}}$
\vskip0.5cm
This work is supported by the SERC, the EC Human Capital and
Mobility Programme and the MOD.
It has benefited from useful conversations with
V. Petrashov, V.C. Hui and F. Sols.

\vskip 2.0truecm
\noindent
$\underline {\hbox {\bf Figure Captions.}}$
\vskip 0.5truecm
\noindent {\bf Figure 1.}
A quantum trombone formed from three one-dimensional wires connected at
a node. This figure shows the arrangement of incoming and outgoing amplitudes
for particles.

\noindent {\bf Figure 2.}
A quantum trombone formed from a two dimensional, tight-binding
lattice of sites. The
superconducting island is coloured black and
located on the vertical arm of the trombone.
The shaded area
represents an insulating portion of the scattering region, while
white area shows the weakly disordered conducting region.
The current flows from
left to right along the horizonatal arm of the T-junction. The external leads
and scatterer
are of width $M=30$ sites. The horizontal arm is 6 sites long.
The horizontal arm and vertical leg are each 2 sites wide.

\noindent {\bf Figure 3.}
Results for the ensemble averaged differential conductance $G(E,L)$
as a function of the position $L$ of the superconducting island,
for $E=0$ (top graph), $E=0.1\gamma$ (middle graph) and $E=0.2\gamma$
 (bottom graph). In each case $\Delta_0=0.2$.
The squares show results for zero applied magnetic field and the
circles for an applied flux of $6\phi_0$ passing through the
normal part of the sample, (ie. the white plus shaded regions of figure 2),
 where $\phi_0=hc/e$.

\noindent {\bf Figure 4.}
As for figure 3, except that $G(E,L)$ is plotted as a function of $E$.
In this figure $\Delta_0 =0.2$ and $L=11$.

\vfil\eject
\noindent
$\underline {\hbox {\bf References.}}$
\vskip 0.5truecm

\item{1.} V.T. Petrashov and V.N. Antonov, Sov. Phys. JETP Lett. {\bf 54}
241, (1991)

\item{2.} V.T. Petrashov, V.N. Antonov and Persson, Physica Scripta,
{\bf T42} 136 (1992),

\item{3.} V.T. Petrashov, V.N. Antonov, P. Delsing and T. Claeson,
Phys. Rev. Lett. {\bf 70} 347 (1993).

\item{4.} G.E. Blonder, M. Tinkham and T.M. Klapwijk, Phys. Rev. B. {\bf 25}
4515 (1982).

\item{5.}  H. van Houten and C.W.J. Beenakker, Physica B. {\bf 175} 187 (1991).

\item{6.}  B.Z. Spivak and D.E. Khmel'nitskii, Sov. Phys. JETP Lett,
{\bf 35} 413 (1982)

\item{7.} C.J. Lambert, J. Phys.: Condens. Matter, {\bf 3} 6579 (1991)

\item{8.} C.J. Lambert, V.C. Hui and S.J. Robinson, J.Phys.: Condens. Matter,
{\bf 5} 4187 (1993)

\item{9.} Y. Takane and H. Ebisawa, J. Phys. Soc. Jap. {\bf 61} 1685 (1992).

\item{10.} C.W.J. Beenakker, R.A. Jalabert and I.K. Marmorkos,
Phys. Rev. {\bf B48} 2811 (1993)

\item{11.} V.C. Hui and C.J. Lambert, Euro. Phys. Lett., {\bf 23} 302 (1993)

\item{12.} C.J. Lambert and V.C. Hui, J. Phys.: Condens. Matter {\bf 5}
L651 (1993)

\item{13.} C.J. Lambert and V.C. Hui, Proc. 20th Int. Conf. on Low Temp.
Phys. (1994) to be published

\item{14.} Y.W. Kwong, K. Lin, P.J. Hakonen, M.S. Isaacson and J.M. Parpia,
Phys. Rev. B {\bf 44} 462 (1991).

\item{15.} P. Santhanam, C.C. Chi, S.J. Wind, M.J. Brady and J.J. Bucchignano,
Phys. Rev. Lett. {\bf 66} 2254 (1991).

\item{16.} A. Kastalasky, A.W. Kleinsasser, L.H. Greene, R. Bhat and
J.P. Harbison, Phys. Rev. Lett. {\bf 67} 3026 (1991)

\item{17.} D.R. Heslinga and T.M. Klapwijk, Phys. Rev. {\bf B47} 5157 (1993)

\item{18.} V.T. Petrashov, V.N. Antonov, S.V. Maksimov, Sov. Phys.
JETP Lett., {\bf 58} 49 (1993)

\item{19.} F. Sols, M. Macucci, U. Ravaioli and K. Hess, J. Appl. Phys.
{\bf 66} 3892 (1988)

\item{20.} F. Sols, M. Macucci, U. Ravaioli and K. Hess, Appl. Phys. Lett.
{\bf 54} 350 (1989)

\item{21} H.U. Baranger, A.D. Stone and D.P. DiVincenzo, Phys. Rev. {\bf B37}
6521 (1988)

\item{22.} W.J. Tomasch, Phys Rev Lett., {\bf 15} 672 (1965)

\item{23.} W.L. Macmillan and J.M. Rowell, Phys Rev Lett., {\bf 15} 453 (1966)

\item{24.} V.T. Petrashov  Private communication.

\end